\documentclass[a4paper,11pt]{article}
\usepackage{jcappub}

\def\imo{i}
\def\re#1{Re(#1)}
\def\im#1{Im(#1)}

\begin{document}
\title{Wormholes versus black holes: quasinormal ringing at early and late times}

\author[\dagger]{R. A. Konoplya}
\emailAdd{roman.konoplya@uni-tuebingen.de}
\affiliation[\dagger]{Theoretical Astrophysics, Eberhard-Karls University of T\"ubingen,\\ T\"ubingen 72076, Germany}

\author[\ddagger]{A. Zhidenko}
\emailAdd{olexandr.zhydenko@ufabc.edu.br}
\affiliation[\ddagger]{Centro de Matem\'atica, Computa\c{c}\~ao e Cogni\c{c}\~ao, Universidade Federal do ABC (UFABC),
Rua Aboli\c{c}\~ao, CEP: 09210-180, Santo Andr\'e, SP, Brazil}

\abstract{
Recently it has been argued that the phantom thin-shell wormholes matched with the Schwarzschild space-time near the Schwarzschild radius ring like Schwarzschild black holes at early times, but differently at late times \cite{Cardoso:2016rao}. Here we consider perturbations of the wormhole which was constructed without thin-shells: the Bronnikov-Ellis wormhole supported by the phantom matter and electromagnetic field. This wormhole solution is known to be stable under specific equation of state of the phantom matter. We show that if one does not use the above thin-shell matching, the wormhole, depending on the values of its parameters, either rings as the black hole \emph{at all times} or rings differently also \emph{at all times}. The wormhole's spectrum, investigated here, posses a number of distinctive features.

In the final part we have considered general properties of scattering around arbitrary rotating traversable wormholes. We have found that symmetric and non-symmetric (with respect to the throat) wormholes are qualitatively different in this respect: First, superradiance is allowed only if for those non-symmetric wormholes for which the asymptotic values of the rotation parameters are different on both sides from the throat. Second, the symmetric wormholes cannot mimic effectively the ringing of a black hole at a few various dominant multipoles at the same time, so that the future observations of various events should easily tell the symmetric wormhole from a black hole.
}
%
%\pacs{04.50.Kd,04.70.Bw,04.30.-w,04.80.Cc}

\maketitle

\section{Introduction}

Gravitational waves emitted by a perturbed black hole are dominated at late times by its proper damped oscillation frequencies called \emph{quasinormal modes} \cite{Konoplya:2011qq,Berti:2009kk,Kokkotas:1999bd}. Owing to the recent detection of a gravitational wave signal, which may be radiation from a merger of two black holes \cite{Abbott:2016blz,TheLIGOScientific:2016src}, there is the revival of interest in study of quasinormal modes of black holes. Quasinormal modes of astrophysical black holes are proper eigenvalues corresponding to the purely ingoing waves at spacial infinity and purely incoming waves at the event horizon. The boundary conditions for the quasinormal modes of black hole and wormholes are the same \cite{Konoplya:2005et}. The wormholes are considered to be exotic objects, because for their existence in the four-dimensional gravity, an unusual, phantom matter is required as a rule. Although, there is one exception: a wormhole in the Einstein-dilaton-Gauss-Bonnet theory, where the Gauss-Bonnet term provides the necessary negative energy density to support the wormhole \cite{Kanti:2011yv}.

Even more problematic is the issue of their stability: usually, the known wormhole solutions are unstable against small perturbations of space-time (see for example \cite{Bronnikov:2012ch,Kokubu:2015spa} and reference therein). Nevertheless, in the past few years there appeared a number of indications that the stable wormholes could exist. Thus, a thin-shell wormhole \cite{Garcia:2011aa} was tested against small purely radial perturbations of the thin-shell \cite{Varela:2013xua} and it was shown that the system is stable against such perturbations, once the matching occurs near the Schwarzschild radius. Radial linear stability of the Einstein-Gauss-Bonnet wormhole (the first one, constructed without any exotic matter), reported in \cite{Kanti:2011yv}, deserves the detailed study of its stability \cite{work-progress}. Stability of the Bronnikov-Ellis wormhole \cite{Bronnikov-Ellis} supported by the phantom matter and electromagnetic field \cite{Shatskiy:2008us} is studied better, though, also not exhaustively. The pressureless phantom matter is unstable both linearly \cite{Bronnikov:2013coa} and, naturally, non-linearly \cite{Sarbach:2010we}. In \cite{Bronnikov:2013coa} it was shown that there are special equations of state which allow for stable configurations of the system against spherically symmetric and non-spherical axial perturbations. Summarizing the stability issue, currently there is no wormhole configuration whose stability was completely confirmed for arbitrary perturbations even in the linear approximation.

Recently, a thin-shell wormhole configuration was used in the following interesting discussion: Two Schwarzschild space-times matched with the phantom thin-shell at the throat was considered in \cite{Cardoso:2016rao}, and the matching is supposed to occur near the Schwarzschild radius. Then, the time-domain computation of the signal coming from a particle near this wormhole, was shown to be very close to that for the Schwarzschild black hole at early times, but different at sufficiently late times. Consequently, it was claimed that very late times of the gravitational ringdown are important to tell the wormhole from the black hole \cite{Cardoso:2016rao} in the observations of the gravitational waves. Notice, that it was first shown in \cite{Damour:2007ap} that the thin-shell wormhole can mimic various properties of the Schwarzschild black hole, in particular, its quasinormal ringing.

Here we shall show that once one does not use the special matching near the Schwarzschild radius and use the configuration constructed without thin-shells instead, an opposite situation occurs: a wormhole may ring \emph{like a black hole at all times or differently also at all times}, depending on the values of the wormhole and black-hole parameters.  For this purpose we shall consider, as an illustration, the above mentioned Bronnikov-Ellis wormholes supported by the phantom matter and electromagnetic field. We shall further consider some generalization of this wormhole for the case of non-zero rotation and some general properties of traversable rotating wormholes. In particular, we find that super-radiance is allowed only for wormholes which are not symmetric with respect to their throats and only under certain exotic assumptions. Then we show that the wormhole symmetric with respect to its throat can mimic effectively the ring-down of a black hole only at a chosen multipole number, but not various modes corresponding to different multipole numbers at the same time.

The paper is organized as follows: Sec.~\ref{sec:perturb} summarizes briefly the perturbation equations for the non-rotating wormhole, including the obtained effective potentials. Sec.~\ref{sec:QNM} is devoted to the time-domain and WKB analysis of the quasinormal spectra of the Bronnikov-Ellis wormholes. Sec.~\ref{sec:wormholeringing} considers ringing of generalized traversable wormholes. Sec.~\ref{sec:wormholesuperradiance} is devoted to general analysis of super-radiance for arbitrary rotating traversable wormholes. In the Conclusion (Sec.~\ref{sec:conclusions}) we discuss the obtained results and future aims.

\section{Perturbations of the Bronnikov-Ellis wormholes.}\label{sec:perturb}

We consider a simple model of a zero-mass wormhole, given by \cite{Bronnikov:2013coa}
\begin{equation}\label{chargedwormhole}
ds^2=-dt^2+dx^2+(x^2+q^2)(d\theta^2+\sin^2\theta d\phi^2),
\end{equation}
where the parameter $q$  characterizes the electric or magnetic field in Wheeler's geometrodynamics \cite{Shatskiy:2008us}. The phantom matter is supposed to have a negative energy density:
\begin{equation}
\epsilon_{DE} = - \frac{q^2}{ 4 \pi (x^2+q^2)^2}.
\end{equation}

I. Novikov, K. Bronnikov, and others showed that under the special choice of the equation of state for the phantom matter, the above wormhole is linearly stable against spherically symmetric perturbations of polar type and arbitrary axial perturbations \cite{Novikov:2012uj,Bronnikov:2013coa}. This equation of state implies that initially pressureless phantom matter acquires non-zero pressure proportion to the perturbations of its density. As currently there is no any strictly pre-viewed equation of state for the phantom matter, this assumption in \cite{Novikov:2012uj,Bronnikov:2013coa} appears valid. The axial perturbations satisfy the wavelike equations \cite{Bronnikov:2012ch}
\begin{equation}\label{wave-like}
  \begin{array}{rcl}
    \dfrac{\partial^2 H_1}{\partial x^2}-\dfrac{\partial^2 H_1}{\partial t^2}&=&V_{11}(x)H_1+V_{12}(x)H_2, \\\mbox{}\\
    \dfrac{\partial^2 H_2}{\partial x^2}-\dfrac{\partial^2 H_2}{\partial t^2}&=&V_{22}(x)H_2+V_{21}(x)H_1,
  \end{array}
\end{equation}
where $H_1$ and $H_2$ define the amplitude of electromagnetic and gravitational perturbations respectively and the potentials have the form
\begin{eqnarray}
V_{11}(x)&=&\frac{\ell(\ell+1)}{x^2+q^2}+\frac{4q^2}{(x^2+q^2)^2},\nonumber\\
V_{22}(x)&=&\frac{\ell(\ell+1)}{x^2+q^2}-\frac{3q^2}{(x^2+q^2)^2},\label{wormholepotentials}\\
V_{12}(x)&=&V_{21}(x)=2q\sqrt{\frac{(\ell-1)(\ell+2)}{(x^2+q^2)^3}},\nonumber
\end{eqnarray}
where $\ell=2,3,4,\ldots$ is the multipole number.

Although the background static configuration assumes that the phantom matter can be described effectively as the pressureless dust, the s-mode perturbations in their general form lead not only to the variation of the energy density, but also to a non-zero contribution to the pressure \cite{Novikov:2012uj}. The arbitrary linear spherically-symmetric perturbations of (\ref{chargedwormhole}) allow for sound waves in the phantom medium, depending on a function
$$h(x)=\frac{\partial p}{\partial \rho},$$
which is square of the speed of sound. The particular choice $h(x)=0$ corresponds to the pressureless matter, imposing an additional constrain to the dynamics of the phantom matter and apparently leading to the instability \cite{Bronnikov:2013coa}.

The phantom matter considered in \cite{Novikov:2012uj,Bronnikov:2013coa} is an exotic scalar field. By now there is no pre-viewed properties of such a field, which would guarantee that once we can effectively describe the static configuration of the phantom matter in equilibrium by the pressureless dust, the time-dependent perturbed configuration could also be modeled by the pressureless dust. That is, we are not guaranteed that the static pressureless configuration is not the particular case of a more general state of phantom matter with non-zero pressure. This freedom was used by authors of \cite{Novikov:2012uj,Bronnikov:2013coa} in order to claim the stability of such a system.

The perturbation equation can be reduced to the following wave-like form \cite{Novikov:2012uj}
\begin{equation}\label{s-mode-wave-like}
\frac{\partial^2\eta}{\partial z^2}-\frac{\partial^2\eta}{\partial t^2}=U\eta,
\end{equation}
where
$$U=\frac{12h(x)x^2+3q^2h(x)-3h'(x)x(x^2+q^2)+q^2}{(x^2+q^2)^2},$$
with the function $h(x)$, defined as
\begin{equation}\label{hchoice}
h(x)=h_0\frac{x^4}{(x^2+q^2)^2},
\end{equation}
and the tortoise coordinate $z$ with respect to the sound waves is related to $x$ through
$$x^2-\sqrt{h_0}zx-1=0,$$
spanning one of the branches, either $x\geq0$ or $x\leq0$. Hereafter we assume that the speed of sound cannot exceed the speed of light, hence, $h(x)\leq h_0\leq1$.

With these master equations at hand we can further exploit the time-domain integration in order to find quasinormal modes and ringing profiles of the above wormhole.

\section{Quasinormal ringing of the Bronnikov-Ellis wormholes}\label{sec:QNM}

We shall use here the discretization scheme proposed by Gundlach, Price, and Pullin \cite{Gundlach:1993tp}. This method was used for calculation of quasinormal modes in a great number of works  \cite{Konoplya:2011qq}. Comparisons of the time-domain numerical data with the accurate frequency-domain calculations show excellent agreement not only in cases when a black hole is stable, but also near the onset of instability (see, for example, \cite{Konoplya:2014lha}).  Rewriting the master equation (\ref{s-mode-wave-like}) in terms of the light-cone coordinates $du=dt-dz$ and $dv=dt+dz$, one finds
\begin{equation}\label{timedomain}
4\frac{\partial^2\eta}{\partial u\partial v}=-U\eta.
\end{equation}

The discretization scheme has the following form
\begin{eqnarray}
\eta(N)&=&\eta(W)+\eta(E)-\eta(S)\label{ci1}%\\\nonumber&&
-\frac{\Delta^2}{8}U(S)\left[\eta(W)+\eta(E)\right]+\mathcal{O}(\Delta^4),
\end{eqnarray}
where $N$, $W$, $E$ and $S$ are the points of a square in a grid with step $\Delta$ in the discretized $u$-$v$ plane: $S=(u,v)$, $W=(u+\Delta,v)$, $E=(u,v+\Delta)$ and $N=(u+\Delta,v+\Delta)$. With the initial data specified on two null-surfaces $u = u_0$ and $v = v_0$ we are able to find values of the function $\Psi$ at each of the points of the grid. Since quasinormal modes and the asymptotical behavior of perturbations do not depend on initial conditions (as confirmed by several numerical simulations), we shall consider the Gaussian wave initial data on $v$-axes (see \cite{Konoplya:2011qq} for more details).

For axial-type perturbations the equations are chained as in (\ref{wave-like}), therefore, after introducing the appropriate light-cone coordinates, $du=dt-dx$ and $dv=dt+dx$, we obtain equation, similar to (\ref{timedomain}), but with the effective potential given in the matrix form
\begin{equation}
4\frac{\partial^2}{\partial u\partial v}\left(\begin{array}{c} H_1\\ H_2\\\end{array}\right) =-\left(\begin{array}{cc} V_{11} & V_{12}\\ V_{21} & V_{22}\\\end{array}\right)\left(\begin{array}{c} H_1\\ H_2\\\end{array}\right).
\end{equation}
Applying the same scheme (\ref{ci1}), were the multiplication is replaced by the corresponding matrix product, we obtain time-domain profiles for both electromagnetic and gravitational perturbations, $H_1$ and $H_2$.

\begin{table}
\centering
\begin{tabular}{|c|c|c|c|}
\hline
$\omega\cdot q$&$n=0$&$n=1$&$n=2$\\
\hline
$\ell=2$&$1.24624-0.19237\imo$&$3.353-0.572\imo$&$1.1-1.5\imo$\\
$\ell=3$&$2.28473-0.31947\imo$&$4.379-0.552\imo$&$1.6-1.8\imo$\\
$\ell=4$&$3.32774-0.38457\imo$&$5.398-0.541\imo$&$2.3-1.9\imo$\\
$\ell=5$&$4.36062-0.41832\imo$&$6.412-0.535\imo$&\\
\hline
\end{tabular}

\caption{Dominant axial modes (in units of $q$) of the charged wormhole (\ref{chargedwormhole}) obtained with the help of the convergent time-domain integration.}\label{tabl:chargedwormhole}
\end{table}

\begin{figure}\centering
\includegraphics[width=\textwidth]{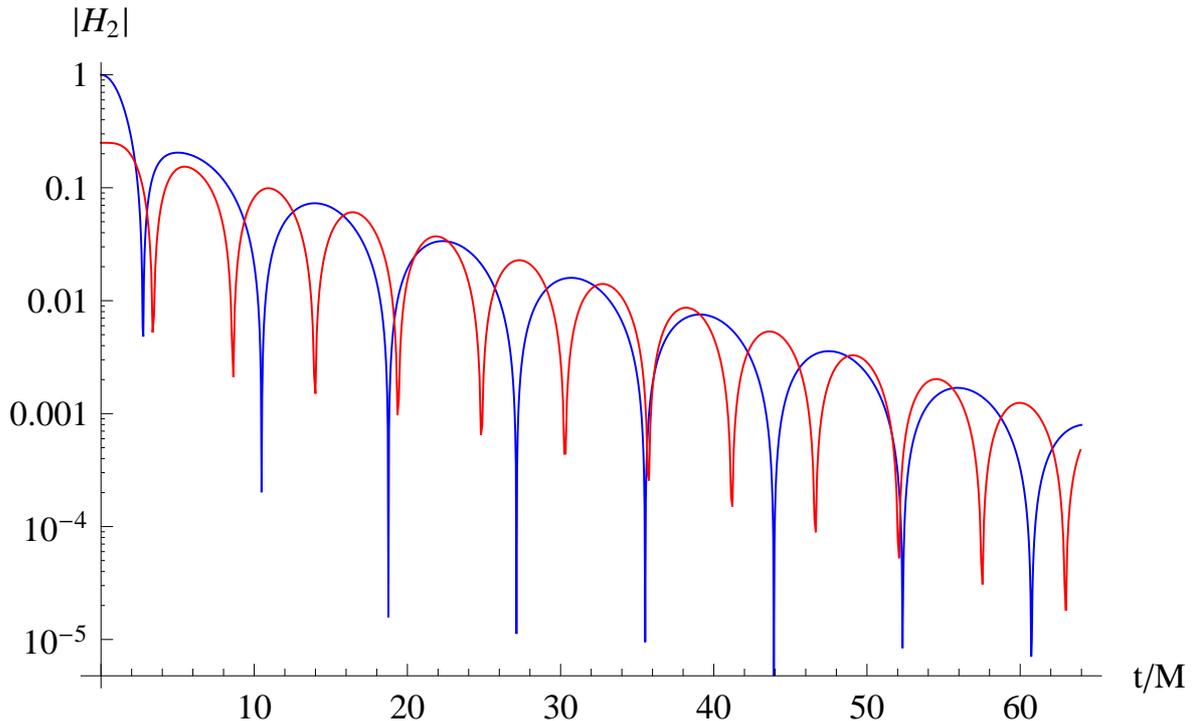}
\caption{Time-domain profiles for the $\ell =2$ axial gravitational perturbations of the Schwarzschild black hole of mass $M$ (blue) and the charged wormhole (\ref{chargedwormhole}) with $q=2.16M$ (red), which has the same decay rate but higher oscillation frequency.}\label{fig:difprofiles}
\end{figure}

The $\ell=2$ ($n=0$) mode $\omega q=1.246-0.192\imo$ (see Table~\ref{tabl:chargedwormhole}) of the Bronnikov-Ellis wormhole has different quality factor $\sim \re\omega/\im\omega$ from that of the Schwarzschild black hole for which $\omega M=0.374-0.089\imo$. This means that one can always tell such a wormhole from a spherically symmetric black hole in General Relativity, even if the corresponding parameters, $q$ and $M$, are unknown (see Fig.~\ref{fig:difprofiles}).

\begin{figure}\centering
\includegraphics[width=\textwidth]{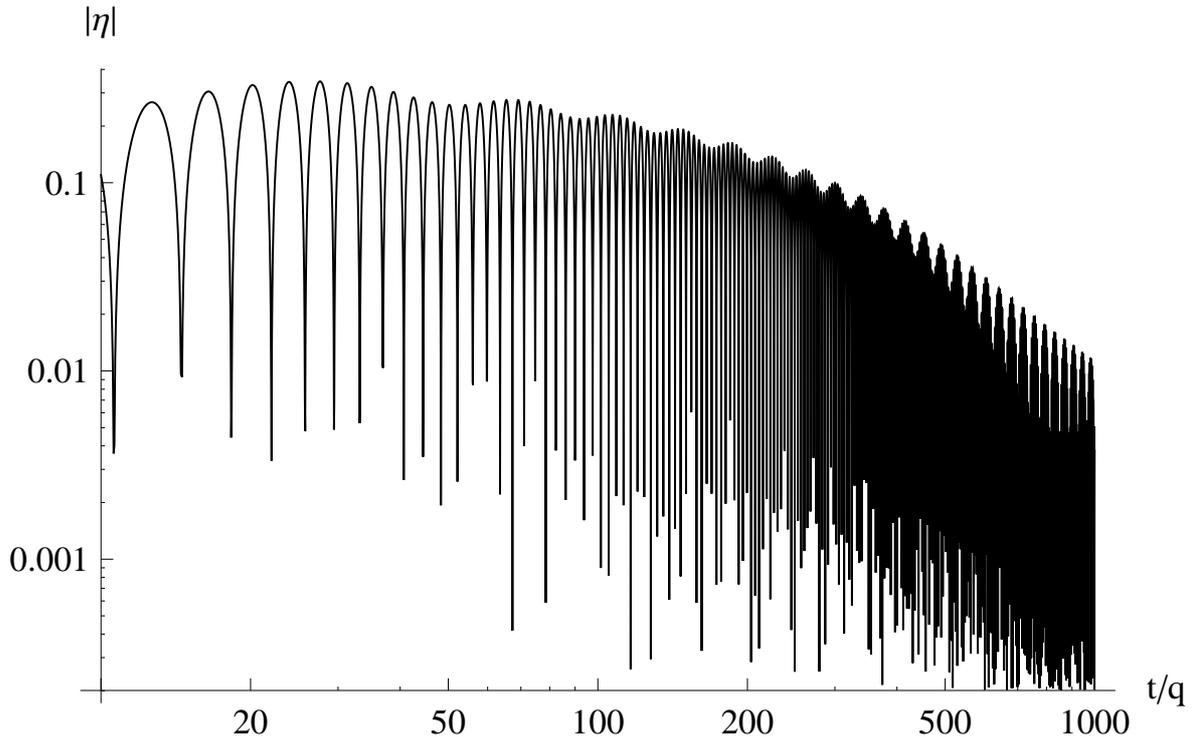}
\caption{The time-domain profile for the s-mode ($h_0=1$) in logarithmic scale. At the asymptotically late time we observe typical massive tales with the effective mass $1/q$. The speed of sound in the phantom dust coincides in this limiting example with the speed of light in order to show that quasi-resonances cannot be achieved.}\label{fig:smode}
\end{figure}

\begin{table}
\centering
\begin{tabular}{|c|c|}
\hline
$h_0$&$n=0$\\
\hline
$0.3$&$0.654-0.061\imo$\\
$0.4$&$0.7011-0.0402\imo$\\
$0.5$&$0.73602-0.02679\imo$\\
$0.6$&$0.76332-0.01814\imo$\\
$0.7$&$0.78517-0.01250\imo$\\
$0.8$&$0.80304-0.00876\imo$\\
$0.9$&$0.81787-0.00625\imo$\\
$1.0$&$0.83035-0.00454\imo$\\
\hline
\end{tabular}
\caption{Dominant s-modes (in units of $q$) which stipulate the perturbations of the phantom matter in the Bronnikov-Ellis wormhole.}\label{tabl:smode}
\end{table}

The particular choice of the equation of state for the phantom matter, made in \cite{Novikov:2012uj}, leads to the non-zero, constant value of the effective potential for the spherically symmetric perturbations at the throat $z= -\infty$. This non-zero asymptotic of the s-wave potential at $z = - \infty$ makes the wave equation similar to that of the massive fields with $\mu^2$-asymptotic of the effective potential. As a result (see Fig.~\ref{fig:smode}), the quasinormal modes and the time-domain profiles look similar to those for massive fields in the background of black holes or of massless fields in the background of black strings \cite{MassiveQNM}. Then, the natural question is whether through varying of the only parameter of the equations of state, the speed of sound at infinity $h_{0}$, one can obtain the infinitely long living modes called quasi-resonances \cite{Quasi-resonance}? When increasing the speed of sound at infinity, we noticed that the damping rate diminishes (see Table~\ref{tabl:smode}), yet, even if $h_{0}$ coincides with the speed of light, the quasi-resonances cannot be reached as it is shown on Fig.~\ref{fig:smode}. However, the choice (\ref{hchoice}) might be not the only one, leading to a stable configuration. Therefore, it is not excluded that other models for the phantom matter may allow for quasi-resonances.

Thus, in the this section we have considered the wormhole configuration which rings differently from a black hole both at the early and asymptotically late times. In the next section, by performing some generalization of the above Bronnikov-Ellis model, we shall show that it is possible to choose such parameters of the wormhole that the latter will also ring as the Schwarzschild/Kerr black hole at early and late times.

\section{Examples of wormholes which ring like the Schwarzschild/Kerr black holes}\label{sec:wormholeringing}

Here we shall consider a more general, traversable wormhole with a few independent parameters in order to illustrate another aspect of the wormhole ringing: parameters of a wormhole can be fixed in such a way that the corresponding quasinormal modes of a wormhole will be very close to those of the Schwarzschild/Kerr black hole under certain values of mass and angular momentum.

We rewrite the metric (\ref{chargedwormhole}) in the general form \cite{Konoplya:2010kv}
\begin{eqnarray}\label{generalwormhole}
ds^2&=&-e^{2\Phi(r)}dt^2+\frac{dr^2}{1-\dfrac{b(r)}{r}}%\\\nonumber&&
+r^2(d\theta^2+\sin^2\theta (d\phi-\bar{\omega}(r) dt)^2),
\end{eqnarray}
by introducing a radial coordinate $r^2=x^2+q^2$. Then, one has
\begin{equation}
e^{2\Phi(r)}=1, \qquad b(r)=\frac{q^2}{r}, \qquad \bar{\omega}(r)=0,
\end{equation}
where $\Phi(r)$ and $b(r)$ are the red-shift and shape functions.

Here, for purely illustrative purpose, we shall consider a more general spherically symmetric Lorentzian traversable wormhole with a throat at $r=q$ by introducing
\begin{equation}\label{massiveqormhole}
e^{2\Phi(r)}=1-\frac{2\mu}{r},\qquad b(r)=\frac{q^2+2\mu(r-q)}{r}.
\end{equation}

The wormhole (\ref{massiveqormhole}) has a number of interesting for us features:
\begin{itemize}
\item It has the same post-Newtonian behaviour as a point-like particle of mass $\mu$,

 \item It has a throat at $q\geq2\mu\geq0$ such that in the limit of $q=2\mu$ it coincides with the Schwarzschild black hole. The similar limit occurs with the thin-shell black hole when the throat is approaching the Schwarzschild radius \cite{Cardoso:2016rao}, though here we do not use any phantom shells.

\item  When $\mu=0$ we have the massless wormhole (\ref{chargedwormhole}) found by Ellis and Bronnikov.
\end{itemize}

Using the standard ansatz
$$\Psi(t,r,\theta,\phi)=\frac{H_0(r)}{r}Y_{\ell,m}(\theta,\phi)e^{-\imo \omega t},$$
Klein-Gordon equation for a test scalar field in this background can be reduced to the wavelike equation \cite{Konoplya:2010kv}
\begin{equation}\label{scalar-rotation}
H_0''(x)+(\omega+m\bar{\omega}(x))^2H_0(x)=V_0 H_0(x),
\end{equation}
where $m$ is the integer azimuthal number, $x$ is the tortoise coordinate with two branches (positive and negative) defined as
\begin{equation}
x(r)=\pm\intop_q^r\frac{dr}{e^{\Phi(r)}\sqrt{1-\frac{b(r)}{r}}},
\end{equation}
and the effective potential is
$$V_0=e^{2\Phi(r)}\frac{\ell(\ell+1)}{r^2}+\frac{1}{r}\frac{d}{dx}e^{\Phi(r)}\sqrt{1-\frac{b(r)}{r}}$$
with the multipole number $\ell=|m|,|m|+1,|m|+2,\ldots$.

\begin{figure*}\centering
%\resizebox{\linewidth}{!}{\includegraphics*{potentials.eps}\includegraphics*{sameprofiles.eps}}
\includegraphics[width=.5\textwidth]{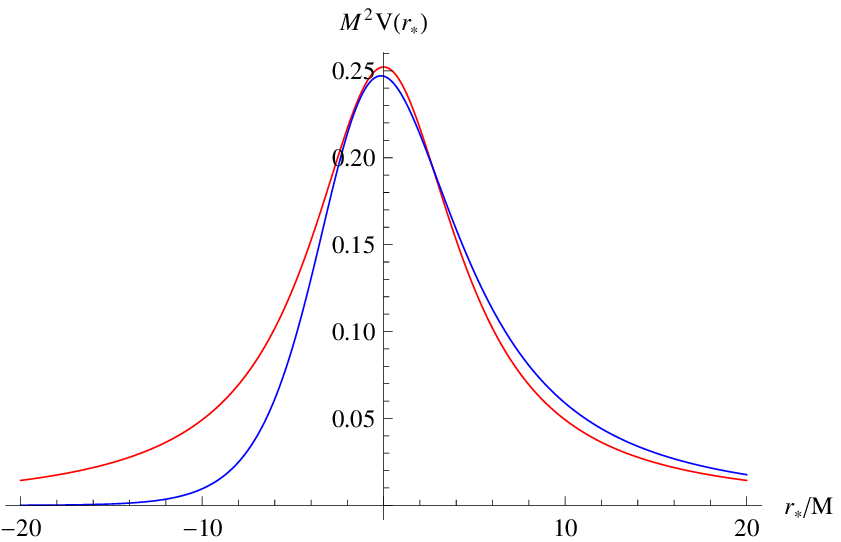}\hfill\includegraphics[width=.5\textwidth]{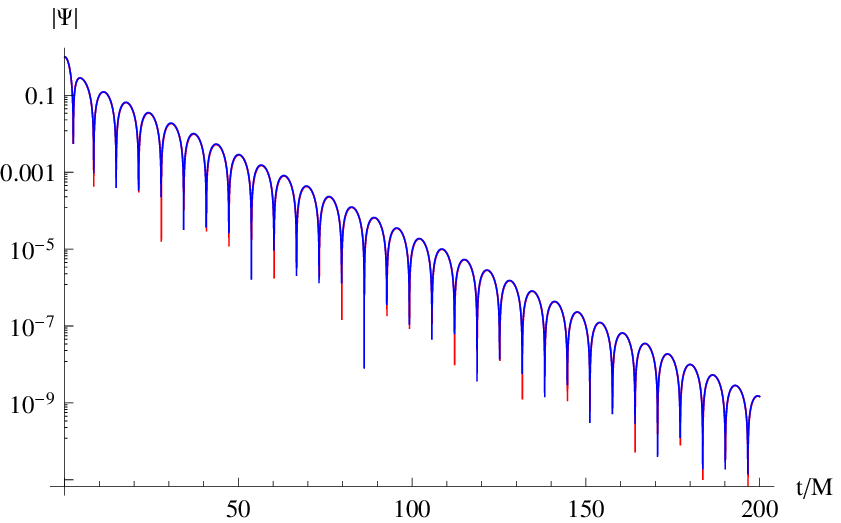}
\caption{The effective potentials (left panel) and time-domain profiles (right panel) for the test scalar field in the background of the Schwarzschild black hole of mass $M$ (blue) and the wormhole with $\mu=M/25$ and $q=5.225M$.}\label{fig:sameprofiles}
\end{figure*}

Since $V_0$ is a function of $r$, due the definition of the tortoise coordinate and, unlike in the Schwarzschild background, we have $V_0(-x)=V_0(x)$. Despite the different behaviour of the effective potential comparing to that of the black hole, one can find such parameters of the wormhole and the Schwarzschild black hole that the $\ell=2$ time-domain profiles  \emph{are the same at all times} (see Fig.~\ref{fig:sameprofiles}). As the parameters of the source emitting the signal GW150914 are currently known with enormous uncertainty of  tens of percents \cite{TheLIGOScientific:2016src}, one has this freedom of arguing, matching signals of a black hole and wormhole with different parameters. Within the black hole concept, the above uncertainty leads also to considerable freedom for deviations from the Kerr geometry, that is, to a window for alternative theories of gravity \cite{Konoplya:2016pmh,Yunes:2016jcc}.

Further generalization of the wormhole can be done by introducing a rotation parameter $a$ into the general wormhole ansatz (\ref{generalwormhole}) as follows
\begin{equation}
\bar{\omega}=\frac{2\mu a}{r^3}\,.
\end{equation}

Application of the WKB formula \cite{WKB}
\begin{equation}
\omega^2 = Q (\ell, m, \omega, M, a, q),
\end{equation}
where the explicit form of the operator $Q$ depends on the order of the WKB series, allows one to find  the dominant quasinormal modes with reasonable accuracy.

\begin{table}
\centering
\begin{tabular}{|c|c|c|}
  \hline
  $\ell$, m & Kerr & rotating wormhole \\
  \hline
  $\ell=m=2$ & 0.635-0.090i & 0.634-0.089i \\
  \hline
  $\ell=m=3$ & 0.904-0.0894i & 0.853-0.0916i \\
  \hline
\end{tabular}

\caption{Dominant quasinormal modes for $\ell=m=2$ for the Kerr black hole ($M=1$ and $a=0.65M$) and the rotating wormhole ($\mu=0.55$, $a=0.65\mu$, $q=3.25$)}\label{tabl:rotatingwormhole}
\end{table}

We can see that the quasinormal frequencies of the Kerr black hole with angular momentum $a=0.65M$ are quite close to those of the rotating wormhole when $\mu=0.55$, $a=0.65\mu$, $q=3.25$. The difference for the $\ell=m=2$ mode is less than the announced by LIGO 3\% accuracy for the observed quasinormal frequency. At the same time $\ell=m=3$ is quite different (see Table~\ref{tabl:rotatingwormhole}). This means that expected observations of more events of gravitational wave emission, say, when two black holes are approximately of the same mass (so that $\ell=m=3$ is excited) should probably quickly tell a wormhole from a black hole. When two black holes (Kerr and its competitor) are compared in this way, the situation is opposite: one can find such parameters of black holes that not only $\ell=m=2$ mode, but many modes of the spectra will be very close for both black holes \cite{Konoplya:2016pmh}. We could suppose that this difference in modes ``fitting'' between the combination comparisons ``black hole vs alternative black hole'' and ``black hole vs wormhole'' is somehow connected with the symmetries of the effective potential mentioned above: the wormhole potential is symmetric with respect to the wormhole throat, thus one apparently cannot fit the potential of a wormhole with the black hole potential everywhere near its maximum. On the contrary, when comparing quasinormal ringing of two black holes such fitting of potentials is certainly possible. Therefore, we expect that when considering an axisymmetric wormhole which is not symmetric with respect to its throat, such fitting wormhole's and black hole's ringings at many modes ($\ell=m=2$, $\ell=m=3$ and others) would be possible.

\section{Scattering around arbitrary rotating traversable wormholes}\label{sec:wormholesuperradiance}

Now we shall consider some general features of waves scattering around such a rotating traversable wormhole, which is not symmetric with respect to its throat. In this section, wishing not to be limited by any particular gravity theory or configuration of fields, we shall develop here an agnostic view on wormholes. Considering in the beginning generic Lorentzian, traversable wormhole, which is not symmetric with respect to its throat, and can even be differentially rotating, we shall show how this general class of geometries could be further constrained by the effect of \emph{superradiance}.

It is clear that unstable metrics cannot exist in nature, and there is no sense in discussing the wave scattering or quasinormal ringing of stable modes on the essentially unstable space-time background. Thereby, although our approach is applicable for these cases as well, we did not consider configurations for which clear indications of instability are known in the literature.

Recently, an interesting rotating non-symmetric wormhole solution supported by the phantom matter has been found in \cite{Kleihaus:2014dla}. Although its non-rotating (and possibly slowly-rotating) regime is unstable \cite{Shinkai:2002gv}, it is not excluded that at some higher rotation the configuration might be stabilized. In that case it would interesting to study the quasinormal ringing of such a wormhole \cite{Kleihaus:2014dla}. Here we can show that as the wormhole solution in \cite{Kleihaus:2014dla} admits different asymptotic values of the rotation parameters on both sides, then the inequality (\ref{asymrotation}) is performed, and the non-symmetric wormhole of \cite{Kleihaus:2014dla} allows for the superradiance.

The metric of a non-symmetric with respect to its throat, rotating, Lorentzian, traversable wormhole can be written as follows \cite{Teo:1998dp}:
\begin{eqnarray}\label{metric11}
ds^2&=&-e^{2\Phi_\pm(r,\theta)}dt^2 + \frac{dr^2}{1 -\dfrac{b_{\pm}(r,\theta)}{r}}
%\\\nonumber&&
+r^2K_{\pm}(r,\theta)(d\theta^2+\sin^2\theta(d\phi-{\bar \omega_{\pm}(r,\theta)} dt)^2),
\end{eqnarray}
where components of the metric tensor, being functions of $r$, and $\theta$, are chosen in such a way that they
are regular on the symmetry axis $\theta = 0, \pi$ and the functions $b_\pm (r, \theta)$ are non-negative ($r =b_\pm(r,\theta)$ corresponds to the throat).
The functions  $\Phi_\pm(r,\theta)$, $b_\pm(r,\theta)$, $K_\pm(r,\theta)$, and ${\bar \omega_{\pm}(r,\theta)}$ must match at the throat, so that the metric coefficients are continuous and differentiable.

The effect of superradiance originally takes place for rotating black holes and conducting cylinders. Black holes spend their rotational energy on amplification of incident waves of perturbation \cite{Zeldovich,Starobinsky}. This amplification can be easily seen from the asymptotic forms of the solutions near the black hole horizon and at spatial infinity, say, for Kerr geometry,
\begin{eqnarray}\label{infty-asym}
\Psi &=& e^{-\imo\omega x} + R e^{+ \imo\omega x}, \quad x \rightarrow \infty,\\
\label{horizon-asym}
\Psi &=& T e^{- \imo(\omega - m \Omega_{h}) x}, \quad x \rightarrow -\infty,
\end{eqnarray}
where $\Omega_{h}$ is the angular velocity of the black hole.

Here $R$ is called the amplitude of the reflected wave, and $T$ is the transmission coefficient. If $|R| > 1$, that is $(m \Omega_h/\omega) > 1$, then, the reflected wave has larger amplitude than the incident one. As the process of super-radiant amplification occurs due-to extraction of rotational energy of the black hole, it happens only for ``co-rotating'' modes with positive values of azimuthal number $m$.

The Klein-Gordon equation in the background given by (\ref{metric11}) does not allow separation of angular variables from the radial part in the most general case. For a rotating wormhole, which is symmetric with respect to its throat, and under the assumption that $\Phi$, $b$, and ${\bar \omega}$ are functions of $r$ only and $K$ is a function of $\theta$, the separation of variable is possible \cite{Kim:2004ph}. It was shown that for this particular case the superradiance is absent \cite{Konoplya:2010kv}.
Here we shall not constrain the behavior by requiring that some of the functions depend only on $r$ and the others -- only on $\theta$ in the whole space. It is sufficient to assume that the dominant contributions to $\Phi_\pm(r,\theta)$, $b_\pm(r,\theta)$, and ${\bar \omega_\pm}(r,\theta)$ at large distance are constants, while the dominant term in $K_\pm(r,\theta)$ far from the throat depends only on $\theta$,
\begin{eqnarray}
\Phi_\pm(r,\theta) &=& {\cal O}\left(\frac{1}{r}\right), \nonumber\\
b_\pm(r,\theta) &=& B_\pm+{\cal O}\left(\frac{1}{r}\right), \nonumber\\
\bar \omega_\pm(r,\theta) &=& \Omega_\pm+{\cal O}\left(\frac{1}{r}\right), \nonumber\\
K_\pm(r,\theta) &=& K_\pm(\theta)+{\cal O}\left(\frac{1}{r}\right). \nonumber
\end{eqnarray}

Using the same ansatz, we find the asymptotic form of the wave-like equation (cf.~Eq.~\ref{scalar-rotation})
$$ \frac{d^2\Psi}{dx^2}=-\left(\omega+m\Omega_\pm+{\cal O}\left(\frac{1}{r}\right)\right)^2\Psi,$$
where the tortoise coordinate at large $r$ is given
$$x = \pm\int \frac{dr}{e^{\Phi_\pm(r,\theta)}\sqrt{1 - \frac{b_\pm(r,\theta)}{r}}}=\pm r+{\cal O}\left(\ln r\right).$$

The tortoise coordinate $x$ (although unknown explicitly in such a general consideration) maps the two regions, $r\geq b_\pm(r,\theta)$, onto $(-\infty,\infty)$ and for the scattering problem we have the following boundary conditions
\begin{eqnarray}\nonumber
\Psi&=&e^{-\imo(\omega+m\Omega_+)x}+R e^{\imo(\omega+m\Omega_+)x}, \qquad x\rightarrow\infty,\\\nonumber
\Psi&=&Te^{-\imo(\omega+m\Omega_-)x}, \qquad\qquad\qquad\qquad x\rightarrow-\infty.
\end{eqnarray}

\begin{figure}\centering
\includegraphics[width=\textwidth]{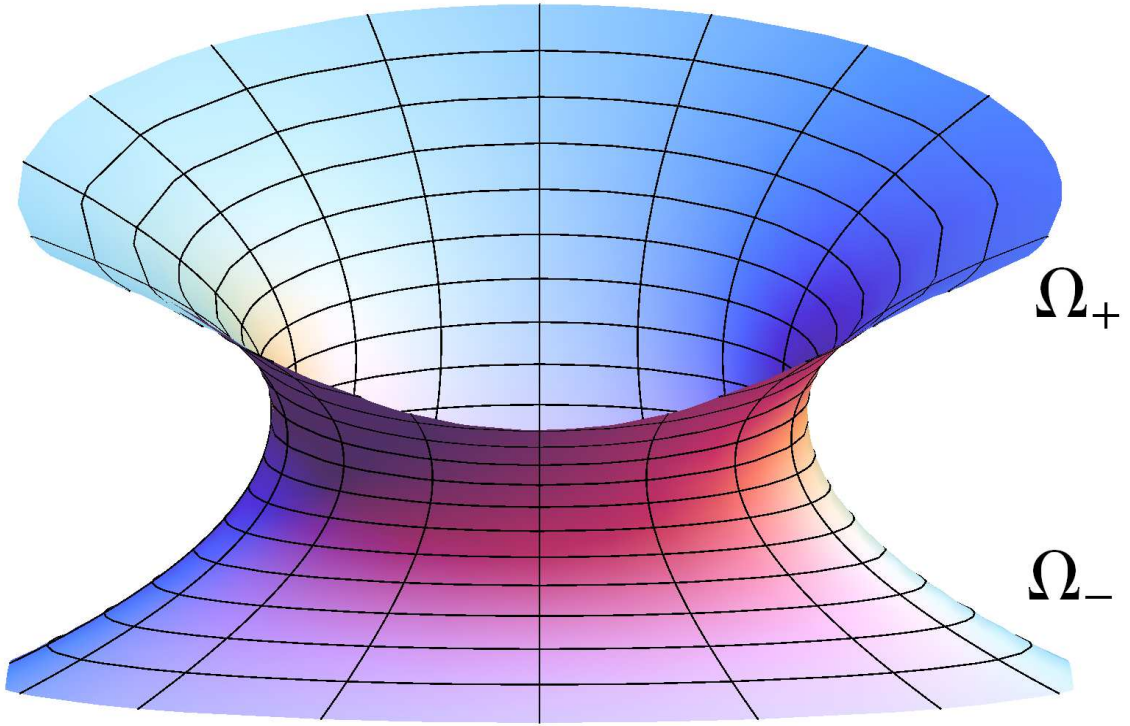}
\caption{Axisymmetric wormhole which is not symmetric with respect to its throat.}\label{wormhole-shape}
\end{figure}

By comparing the Wronskian of the two linearly independent solutions of the wave-like equation $\Psi$ and $\Psi^*$ at the boundaries, we find that
$$|T|^2+|R|^2=1$$ unless
\begin{equation}\label{asymrotation}
\Omega_+\neq \Omega_-,
\end{equation}
that is, unless the rotation parameters as seen from the asymptotically far regions at both sides from the throat are not equal. The latter corresponds to a quite exotic situation of the differentially rotating wormhole or, probably, to a situation when the ``right universe'' rotates relatively to the ``left'' one (see Fig.~\ref{wormhole-shape}). The super-radiance, evidently, should occur until the asymptotic rotation parameters $\Omega_\pm$ become equal. Following the path of exotic models, one could say that if wormholes between various rotating, G\"{o}del-like universes existed, after some time the super-radiance would have prevented difference in the speed of rotation between these universes.

\section{Conclusions}\label{sec:conclusions}

The radiation of a point particle in the vicinity of the phantom thin-shell wormhole matched with Schwarzschild space-time near the Schwarzschild radius leads to the time-domain radiation profile which is different from the Schwarzschild one's only at sufficiently late times \cite{Cardoso:2016rao}. This leads to the conclusion that in order to test the regime strong gravity (a horizon or a wormhole throat) one needs to investigate the very late times of the gravitational ringdown. The existence of such an exotic distribution of the phantom matter near the wormhole's throat, that is, the thin-shell model, as well as the particular matching near the Schwarzschild radius are bold assumptions which should certainly be well justified \cite{matching}.

Here we shave shown that if one does not use the above thin-shell ``tailoring'' and a specific matching near the Schwarzschild radius, then, the opposite conclusion takes place: Varying the shape function of a traversable wormhole we show that its quasinormal ringing can be ``made'' indistinguishable from that of the black hole \emph{at all times}. On the other side, once the wormhole's ringing is different from the black hole's one, it is different also \emph{at all times}. The latter is the case of the Ellis-Bronnikov wormhole configuration supported by the electromagnetic field and the phantom matter, which was constructed without thin-shells. We showed that while $\ell \geq 2$ modes of this wormhole are different from the Schwarzschild ones, the $s-mode$ looks like a perturbation of an effectively massive field. The latter is owing to the special equation of state of the phantom matter.

We have also shown that arbitrary axisymmetric, traversable wormholes allow for superradiance only if they are not symmetric with respect to their throats in such a way that the asymptotic values of the rotation parameters are different on both sides from the throat. We expect that the symmetric wormholes should be different from the non-symmetric ones also in the following aspect: the symmetric wormholes cannot mimic effectively the ringing of a black hole at a few various dominant multipoles at the same time. Therefore, future observations of various events should easily tell the symmetric wormhole from a black hole.

Taking into consideration the above discussed results and supposing that such exotic objects as wormholes might exist, observations in the electromagnetic sector could be the most helpful for telling the wormhole mimicking gravitational ringdown of the black hole.

\begin{acknowledgments}
We would like to acknowledge Kirill Bronnikov, Sergey Solodukhin, and Olivier Sarbach for useful discussion. A.~Z.~was supported by Conselho Nacional de Desenvolvimento Cient\'ifico e Tecnol\'ogico (CNPq).
\end{acknowledgments}

\end{document}